\documentclass[letterpaper]{article}
\usepackage[utf8]{inputenc}
\usepackage{url}
\usepackage{isea}
\usepackage[pdftex]{graphicx}
\usepackage{times}
\usepackage{helvet}
\usepackage{courier}
\usepackage[final]{changes}
\usepackage{caption}
\usepackage{subcaption}
\usepackage{mciteplus}
\usepackage{hyperref}
\usepackage[english]{babel}
\usepackage[numbers]{natbib}

\newcommand{\xcomment}[1]{}

\title{Enable Time-Sensitive Applications in Kubernetes with Container Network Interface Plugin Agnostic Metadata Proxy}
\author{ 
	Ferenc Orosi, Ferenc Fejes \\
	Ericsson Research TrafficLab \\
	Budapest, Hungary \\
	ferenc.\{orosi, fejes\}@ericsson.com\\
	\newline
}

\begin{document} 
\maketitle

\begin{abstract}
Application deployment in cloud environment is dominated by Kubernetes-orchestrated microservices.
Provides a secure environment, networking, storage, isolation, scheduling, and many other abstractions that can be easily extended to meet our needs.
Time-Sensitive Applications (TSAs) have special requirements for compute and network.
Deploying TSAs in Kubernetes is challenging because the networking implemented by Container Network Interface (CNI) plugins is not aware of the traffic characteristic required by Time-Sensitive Network.
Even if a network interface supports TSN features (e.g.: Scheduled Traffic) and a modified CNI plugin is aware of this interface, the pod network isolation built on top of Linux deletes the metadata required for TSN protocols to work with.

We propose \textit{TSN metadata proxy}\footnote{\url{https://github.com/EricssonResearch/tsn-proxy}}, a simple architecture that allows any TSA microservice to use the TSN capabilities of the physical NIC, without any modification.
This architecture is tightly integrated with the Kubernetes networking model, works with popular CNI plugins, and supports services such as ClusterIP, NodePort, or LoadBalancer without additional configuration.
Unlike former proposals, this architecture does not require either bypassing the Linux kernel network stack, direct access to the physical NIC, escalated privileges for the TSA microservice, or even modification of the TSA.
\end{abstract}

\section{Introduction}

High-tech manufacturing, automotive, avionics, defense, professional audio-video and similar domains have deterministic communication requirements.
These are high reliability, no packet loss due to congestion, low latency variation (jitter), and bounded maximum latency.
An application with these network requirements is called a time-sensitive application (TSA).
To ensure proper TSA operation, both compute and network specific requirements have to be fulfilled.

\subsection{Time-Sensitive Networking Overview}
On the network side, Time-Sensitive Networking (TSN) provides Ethernet networks with tools to meet meet the deterministic communication requirements.
The TSN Task Group \cite{tsntg}, part of the IEEE 802 Working Group, is driving the standardization of these protocols and algorithms.
These standards enable vendor-independent TSN solutions and also the implementation of these TSN features in the mainline Linux kernel.
The TSN implementation in the Linux kernel is discussed in \cite{tsn_in_linux}.

A TSN network can be dimensioned and planned for the TSA requirements, so that congestion-induced packet losses do not happen under normal operation.
Some TSN queuing solutions require synchronized notion of time.
As a result, the time slots of different traffic classes and applications are coordinated end-to-end to meet the allocated bandwidth and latency requirements.

\subsection{Time-Sensitive Computing}
On the compute side, operating systems such as specialized RTOSs or general-purpose Linux with real-time scheduling (part of the mainline since version 6.12) are capable of meeting hard real-time deadlines for TSAs.
For the rest of the work, we focus on Linux-based deployment, but details of real-time task scheduling and supporting APIs are out of scope.
Linux provide API for the TSA to express timing (or deadline) and priority requirements to the TSN.
It also provides a common configuration interface for the operator to setup the TSN forwarding in the network interface.

The compute node connects to the network via a TSN-enabled NIC.
Without this, Linux does it's best to emulate TSN features in software, but these may not meet the tight timing requirements, e.g.: software-based time synchronization is orders of magnitude less accurate without PTP \cite{ieee1588} capable NIC.

\label{sec:tsadeploymetn}
\subsection{Time-Sensitive Application Microservice Deployment on Scale}

For large-scale (hundreds and above) deployment of traditional applications as microservices has been one of the main directions of the last decade.
\textit{Kubernetes} \cite{k8s} has become the de-facto orchestration platform for deploying, operating, monitoring, and managing microservices.
The platform is easily extensible, with customizable pod scheduler, network, storage and device management, as well as the API itself with custom resource definitions (CRDs).

However, orchestration of TSAs in Kubernetes is challenging for various reasons, which are discussed in detail in the following chapter and in the related work \cite{fabos, endpoint_arch, tsn_ebpf, k8s_tsn, k8s_tsn_bme}.
The main difficulties can be summarized in the following points:

\begin{enumerate}
	\item Lack of abstractions needed to request real-time scheduling for TSA microservices
	\item Cluster-wide time synchronization with PTP up to the TSA
	\item Centralized policing and packet scheduling of traffic classes based on their cycle times and priorities
	\item Unprivileged access to TSN NIC hardware capabilities from TSA microservices
	\item Provide network security and isolation for the TSA while passing per-packet scheduling metadata to the TSN NIC
	\item Rely on established send/receive primitives (e.g., Linux/BSD socket API) to avoid modification of the TSA
	\item Integration with various Kubernetes network implementations (CNI plugins) and container runtimes
\end{enumerate}

Our work, as well as the related work to which we refer, focuses mainly on items 4.-7.
While the remaining challenges are also the subject of active research, we consider them out of scope.
To cope with challenges 4.-7., we propose a \textit{TSN metadata proxy} (referred as TSN proxy in rest of the paper for simplicity).
This is a small network plugin for Kubernetes, proposed to use in companion with full-featured, industry grade third party network plugins.
TSN proxy does not require modifications in the TSAs, their container images, Kubernetes or the third-party network plugins.
As such it is very easy to integrate into a Kubernetes cluster by applying only a manifest file as usual.

\section{Time-Sensitive Applications with Kubernetes}

Kubernetes is a general framework, it does not distinguish between TSA pods and normal pods.
This does not preclude the orchestration of TSA pods, but data plane and control plane specificities must be taken into account.
In this chapter, we will look at these peculiarities and the challenges that arise from them.

\subsection{Control plane}
Kubernetes CNI plugins are responsible for implementing the network configuration \cite{k8s_cni_plugin, cni_dev}.
Examples of prevalent CNI plugins are \textit{Antrea}, \textit{Calico}, \textit{Cilium}, \textit{CN2}, \textit{Weave} or smaller ones with limited functionality like \textit{Flannel}, \textit{Kindnet}.
Usually a single CNI plugin configures the entire Kubernetes cluster network.
It is not specified how this is done, control plane just assigns an IP address and network namespace to the newly started pod \cite{k8s_net_model}.
\textbf{Challenge: CNI plugin diversity, many implementation strategies may prevent TSA operation.}

How that IP is accessed, or how the pod reaches other pods or the outside world, is entirely up to the CNI plugin.
The plugin can configure this in a kernel-bypassed way e.g. DPDK, AF\_XDP \cite{xdp} or via traditional Linux network configuration (\textit{iproute2} package tools \cite{iproute2}).
The Kubernetes control plane components e.g.: \textit{etcd}, \textit{kube-apiserver}, \textit{kube-scheduler}, \textit{CoreDNS} can only access this IP address.
This is a problem for TSAs that do not use Layer 3 networking, but we assume here only TSAs communicate over IP.
\textbf{Challenge: control plane has limited knowledge of the network.}

The use of multiple plugins is allowed and supported as standard, in which case control plane will run them in order.
In this case, each plugin sees the output of the previous ones in standard JSON format.
It is important that the first CNI plugin receives the IP address from control plane, which it configures for the pod.
Typically, it also creates the pod's (usually virtual) network interface \cite{veth} and performs most of the configuration.
This primary CNI plugin is usually one of the applications mentioned above, and most of the time no other is needed.

There are frameworks that allow the operator to assign multiple network interfaces to a pod at the same time.
This allows multiple CNI plugins to be used together, with each interface accessing a network configured by a different plugin.
We mention this because it is a popular method for TSN network access, and we will describe such proposals later.
This is done by making the primary CNI plugin a meta-plugin that calls additional CNI plugins, for example some of the ones from the list above, or a separate TSN CNI plugin.
Commonly used meta-plugins are Multus \cite{multus} and CNI Genie \cite{cni_genie}.
In this case, the pod will have a managed IP address and interface visible to the control plane, and one or more additional interfaces outside the control plane's scope, with their own IP addressing.

It is important to point out that this is not a standard operation and is not supported by Kubernetes control plane.
Active work has been going on for several years to extend Kubernetes with an API to support this functionality in the Multi-net SIG \cite{multinet_sig}.
The status of the Multi-net SIG is uncertain and based on the user-stories section in the documentation, it represents a significant configuration burden.
\textbf{Challenge: multiple pod interface and multiple CNI plugins not officially supported.}

In our view, TSA pods do not need a separate interface and data plane for TSN communication.
A small addition of network access configured by a primary CNI plugin is sufficient.
What has made this multiple interface the main direction in other proposals, however, is the delivery of TSN metadata to a physical NIC for TSN network scheduling.
Because, as we will see in the next subsection, this is not trivial if we follow the Kubernetes isolation rules.
Bypassing these with a secondary interface and data plane simplifies the situation.
\textbf{Main challenge: enable TSA pods and remain consistent with the Kubernetes control plane.}

\subsection{Data plane}

Kubernetes works with containers that are available in a local or remote registry.
These containers bundle the TSAs with their dependencies.
Importantly, in the industry, these are created as an end product of a CI pipeline, testing them is part of the process.
Therefore, patching or modifying them after the fact is not part of best practice, and may not even be possible.
\textbf{Challenge: access to the TSN services must be transparent to the application.}

The container runtime orchestrated by Kubernetes executes the containerized TSA.
There are several such runtimes, e.g. \textit{podman}, \textit{Docker engine}, \textit{containerd}.
What they have in common is that they implement container isolation using the Linux kernel namespace subsystem.
Similarly, network isolation of containers is implemented using Linux network namespaces through CNI plugins.
The containerized TSA also runs in such a network namespace and does not interfere with the host and other container networks.

The \texttt{struct sk\_buff} (hereafter \texttt{skb}) structure represents the packet in the Linux network stack with its contents and metadata.
This is the most important data structure for the Linux packet processing, used both for ingress and egress directions.
When any application send data through a socket, \texttt{skb} is allocated in the kernel and processed all the way to the network device driver which only free the \texttt{skb} when the data sent to the wire.
The ingress direction is similar, to speed up processing the driver usually takes \texttt{skb}s from a pre-allocated pool for the received data, which only released when the packet dropped or the userspace consumed the received data.

Part of the network isolation is that the metadata of \texttt{skb} is stripped before moving to another namespace.
This is a problem because TSN-aware packet scheduling requires metadata.
The two most important metadata are timestamp and priority (\texttt{skb->tstamp} and \texttt{skb->priority}).
These are set by the TSA through the socket API as \texttt{SO\_PRIORITY} and \texttt{SO\_TXTIME} auxiliary messages (also known as a control messages) to the data being sent.
Linux's TSN queueing-disciplines (Qdiscs) e.g. \texttt{taprio}, \texttt{etf}, \texttt{mqprio} (in case of frame preemption) schedule packets based on this.
These Qdiscs are configured on the node level, not inside the pod.
Therefore when the \texttt{skb} originates from the pod reach them, they already went through the isolation steps.
Also, if the NIC is TSN capable, it can transparently offload these Qdiscs, resulting in very accurate scheduling.
VLAN priority tags are also set based on the \texttt{skb->priority} value.
\textbf{Challenge: Linux namespace isolation remove every metadata including the TSN related ones.}

Metadata deletion is done in the kernel by the \texttt{skb\_scrub\_packet} function \cite{skb_scrub}, and \texttt{skb->priority} is deleted by \texttt{dev\_forward\_skb}.
The latter has not much to do with namespace isolation, it moves the \texttt{skb} sent from one interface to the receive queue of another interface (which is the common scenario in container networking).
The \texttt{skb->mark} metadata is also deleted, but this cannot be used by TSAs because it is used by CNI plugins for their own policies.
It is worth noting that if there is no network namespace crossing, \texttt{skb->mark} and \texttt{skb->tstamp} are not deleted.
This means that during packet encapsulation, these metadata fields can carry TSN information.
However, since the TSA runs in an isolated pod, these metadata fields are also deleted, so they cannot be used.
\textbf{Challenge: there is no option to configure a set of metadatas to exclude from the isolation process.}

\section{Related Work}

There is also academic and industrial research on cloudification of industrial applications.
This includes some that define a more remote, end-to-end architecture and some that look at low-level details.
In this session, we will look at work whose problem statements are similar to ours and to each other.
The focus is on getting metadata for TSN network scheduling to the NIC, not hard real-time scheduling and CPU partitioning.
Related work addresses this in different ways and also discusses some challenges 

FabOS \cite{fabos} is one of the first papers to look at containerization of vPLCs.
vPLCs are specialized TSAs, therefore this use-case fits well in our scope.
It describes the industrial environment very well and proposes an architecture to integrate CUC/CNC into Kubernetes.
The implementation is based on Multus, creating VLAN subinterfaces on the physical NIC, but no further details are provided.
Unfortunately, no configuration examples or open source code is provided, so the operator would have to figure out exactly what is needed to use it.

Oechsle et. al. \cite{endpoint_arch} presents a high-level, end-to-end solution for cloud-native TSAs.
No implementation details are provided, but what is revealed is that a new application API would be required.
The send/receive primitives would be given a flow ID as a parameter to facilitate scheduling.
Unfortunately, this would require modification of existing applications.
It is also not revealed how it would integrate with Kubernetes, or how it would use the TSN hardware capabilities.

Wen et. al. \cite{tsn_ebpf} extends Cilium, a widely used CNI plugin based on eBPF, with TSN traffic filtering.
It redirects TSN traffic to the host NIC using XDP.
From the measurements, the benefits of this are not clear, and it is a kind of kernel bypass, so egress policies or NAT rules are skipped.
There are also no details on how a more complex deployment would work with this solution, nor support for TSN NICs.

Rosa et. al. proposes a Multus-based CNI plugin that provides a TSN network for pods \cite{k8s_tsn}.
The TSN network is thus provided by a kernel bypass solution using the DPDK.
It uses \texttt{LD\_PRELOAD} to route packets to the TSN network, replacing the send/recv primitives.
While this solution improves forwarding latency, the DPDK makes the hardware capabilities of the TSN NIC unusable.
There is no actual Kubernetes integration based on the code (contrary to what the article claims), and Docker requires manual configuration of Linux commands in a cumbersome manner.
However, the implementation description is detailed enough that a dedicated operator could build their own version around the proposal.
But with all things considered, the proposal might not met the requirements of a real-life deployment.

Balla et. al.'s proposal \cite{k8s_tsn_bme} is also a Multus-based solution, but is much closer to a real-world deployment.
The application pods here also use modified send/recv primitives that send the packet and TSN metadata to shared memory.
From there, a monolithic application reads it and delivers it to the physical interface via a MACVLAN interface.
Measurements show that the hardware capabilities of the TSN NIC make the timing of packet delivery very accurate indeed ($\sim$10ns).
This places them in the correct timeslot for IEEE 802.1Qbv, resulting in prioritized TSN traffic over best-effort.
There are some shortcomings in the proposal.
This is also a kernel bypass with shared memory and is built on an alternate network, so application egress and ingress policies are not enforced.
The Kubernetes integration is not detailed, nor is how this would work with servicers (\textit{ClusterIP}, \textit{NodePort}, \textit{LoadBalancer}, etc.).
However, it would be easy to overcome the shared memory kernel bypass by creating the pod MACVLAN interfaces directly through the CNI plugin.

\section{Proposed Architecture}

\begin{figure}[tp]
	\centering
	\includegraphics[width=1\columnwidth]{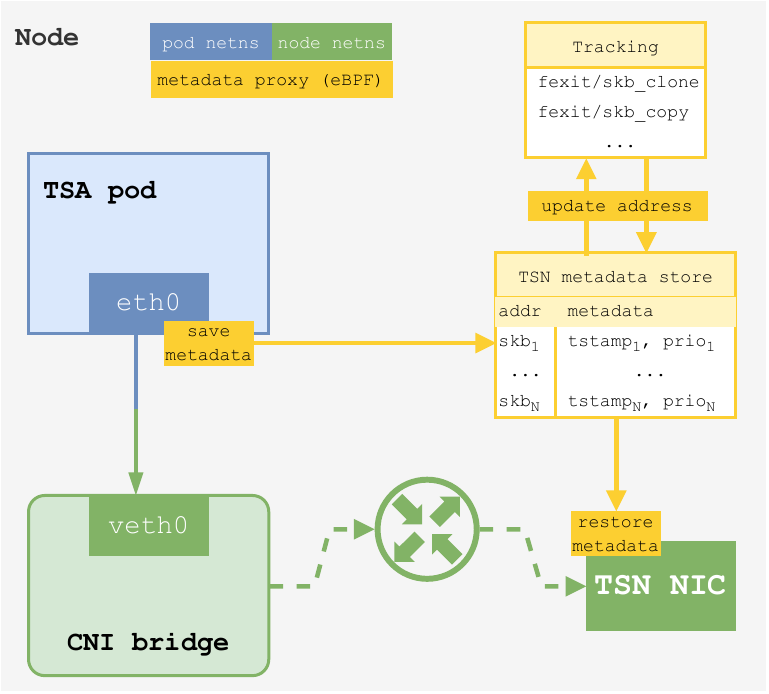}
	\caption{Architecture of the TSN metadata proxy.
		A proxy instance on every node in the cluster setup a storage for the TSN metadata.
		With eBPF it collects the metadata from each packet into the storage and restores it before reaching the TSN NIC.
		A tracking eBPF program update the memory address of the packet if changes during the forwarding.}
	\label{fig:tsnproxy}
\end{figure}

Prior work as discussed above successfully address some of the challenges but fails in others.
Some solution required to used as a primary CNI plugin (or sub-plugin of Multus) even though they are limited in functionalities.
While modifying a primary CNI plugin (as Cilium in \cite{tsn_ebpf}) to support TSN is a good direction, it has to be part of the plugin's mainline code, because these frequently updated with new functions, bugfixes and security patches.
Some proposals leverage kernel-bypass for reduced latencies.
This not only prevent using the NIC TSN functionalities, but often comes with custom built kernels and modules as well.
TSA modification or custom socket API primitive hooks also part of some proposals.
\textbf{Our proposal does not require TSA modification or socket API hooks.
Also, we do believe instead of changing the primary CNI plugin, we have to leverage them and equip those with TSA support in a way not tied to a particular plugin, instead works with most of them.
Lastly, our proposal does not require to bypass the whole Linux network stack, which has compatibility, security and observability advantages.}

The main goal is to overcome these challenges with TSN proxy which is tightly integrated into Kubernetes through standard interfaces.
In other words, a standard \texttt{kubectl apply} would do all the necessary configuration without manual intervention.
This allows scalability of the proposal to multi-node and pod environments.

The tasks can be divided into two parts, the first is the preparation of the node for the TSN proxy, the second is the network configuration of the pods.
The resulting environment summarized on Figure \ref{fig:tsnproxy}.

\subsection{Steps of the node initialization}

\begin{enumerate}
	\item The TSN proxy \textit{DaemonSet} creates an eBPF \cite{ebpf} hashmap for the metadata on each node. The \textit{DaemonSet} ensure the container in the manifest are deployed on each node in the cluster in one instance. The eBPF hashmap will be a node-wide accessible storage of the TSN metadata, available from all network namespace including the node's default namespace.
	\item Installs the TSN proxy binaries on the node, including the TSN proxy CNI plugin script executed when a new pod started, the eBPF programs to save, restore metadata and track the packet as well as a statically built \texttt{bpftool} in order to allow execution on nodes where it is not available or old version installed.
	\item Adds the TSN proxy to the list of CNI plugins so that it runs last when the primary CNI plugin is done. E.g.: if the primary CNI plugin is Flannel, it parse the \texttt{.conflist} JSON in \texttt{/etc/cni/net.d} folder and insert the TSN proxy CNI plugin after Flannel in the list.
	\item On the interface specified in the manifest (e.g.: the TSN NIC), it loads a \texttt{tc egress} program responsible for restoring the metadata. This is done in the node's default namespace. The \texttt{tc egress} eBPF programs executed on each packet (\texttt{skb}) right before they passed to the Qdisc.
	\item Installs an eBPF \texttt{fexit} probe to track \texttt{skb} copies. The probe can execute arbitrary eBPF program which has access to the function arguments, local variables, and the return value of the inspected function (more on that later).
	\item Starts a garbage collector that runs every few seconds, which responsible to remove old entries from the metadata hashmap. Why it is necessary discussed later.
\end{enumerate}

\subsection{Steps of the pod initialization}

\begin{enumerate}
	\item When the pod is created, the \texttt{kubelet} creates the namespaces, including the network namespace for the pod.
	\item The primary CNI plugin creates the interface to the pod, this is usually a connected pair of \texttt{veth}s \cite{veth}, one instance of which is in the pod and the other in the host network namespace.
	\item The CNI plugins are executed in order as defined in the CNI config, after the primary plugin the TSN proxy setup is executed. This assigns a \texttt{tc egress} eBPF program to the pod's \texttt{veth} interface (this is inside the pod's network namespace), which does the storing of the metadata. Important to note here, the \texttt{skb} tracking and metadata restore programs already configured during the node initialization detailed before.
\end{enumerate}

The metadata store uses the eBPF \texttt{tc egress} entry point, not the one where the isolation and metadata deletion is done, \texttt{skb\_scrub\_packet} \cite{skb_scrub}.
This is because it is targeted only at pods that are running TSA (the node may have pods running where it is unnecessary because it is not running TSA).
Important to emphasize the full payload of the packets are not copied or modified by the proxy.
Only the metadata fields proxified, and this is a very lightweight operation because those are 12 bytes in summary (4 bytes priority, 8 bytes timestamp).

\subsection{Tracking the packets through the node}

After leaving the pod, the fate of the packet depends on the primary CNI plugin configuration.
It may simply be bridged or routed to another pod within the node and never reach the physical network.
In a more realistic setup, it will be sent through the physical NIC to the TSN.
But before that, it may be encapsulated in a tunnel, dropped by an egress policy, or modified by NAT rules.
In some cases, the original \texttt{skb} structure may be reallocated, causing the change in its memory address.
Without the address, there is no way to look up the metadata for the packet.
To overcome this, we need to track the packet inside the node.
More specifically, we need to identify and track when the memory address of the \texttt{skb} representing the packet changes.
If the memory address changes, the proxy must update it as well, adding a new entry to the BPF map with the new address and the known metadata.

For tracking, two alternative options are possible.
One is the method used by the \textit{pwru} packet tracking software \cite{pwru}, which is also used by the TSN proxy.
It puts a \texttt{fexit} eBPF probe on the \texttt{skb\_clone} kernel function, which copies \texttt{skb} structure where necessary.
The \texttt{fexit} probe is similar to \texttt{kretprobe}, but it also works with type information so it is more convenient to use.
This probe is injected and executed right after the inspected kernel function returns.
This is enough information to implement the tracking functionality.

When the \texttt{skb} is copied by the \texttt{skb\_clone}, the eBPF probe checks if the metadata hashmap contains a source \texttt{skb} address.
If so, it will re-add the metadata using the new \texttt{skb} address as a key.
This way, when the metadata recovery program is run, which already sees the new \texttt{skb} address, it can successfully assign the original metadata.

Another, simpler tracking method is to use the \texttt{skb->data} address as the key instead of the \texttt{skb} address.
The new \texttt{skb->data} field after \texttt{skb\_clone} is also equal to the old \texttt{data} field.
In practice this has also proved sufficient, but more extensive testing would be needed to see if it is robust enough.
This solution does not require a separate \texttt{fexit} probe package tracker, only the metadata saver and restorer.

Some packets sent by the TSA might conceivably be dropped and not get out of the node.
This could be caused by a firewall rule or configuration problem.
Therefore, there is a periodic garbage collection that looks through the hashmap and deletes anything with a timestamp that is too old.
When the packet reaches the TSN NIC and the metadata is restored, it is also deleted from the hashmap.

\section{Operation Example}

\begin{figure}[htbp]
	\centering
	\begin{subfigure}[b]{0.5\textwidth}
		\centering
		\includegraphics[width=0.9\textwidth]{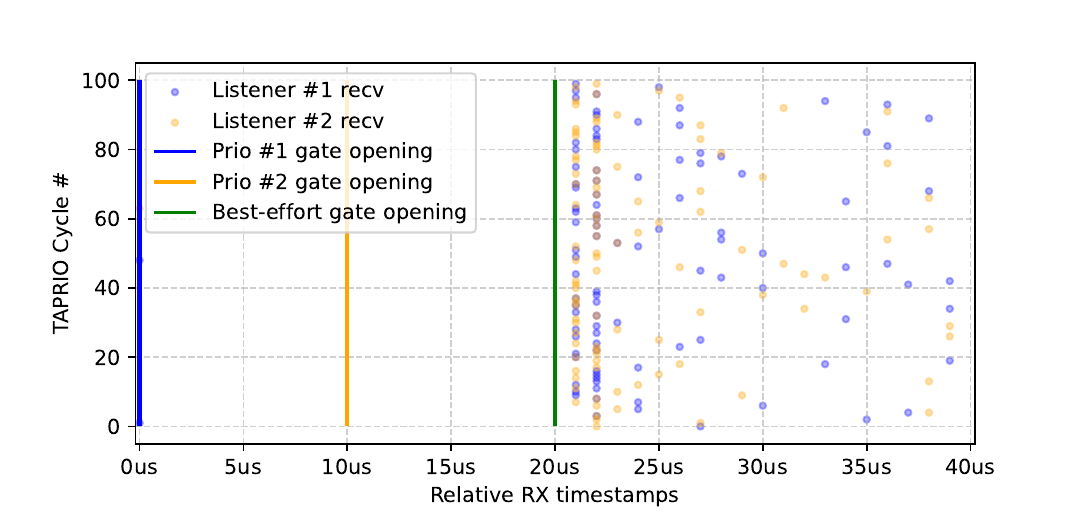}
		\caption{Without TSN metadata proxy}
		\label{fig:noproxy}
	\end{subfigure}
	
	\begin{subfigure}[b]{0.5\textwidth}
		\centering
		\includegraphics[width=0.9\textwidth]{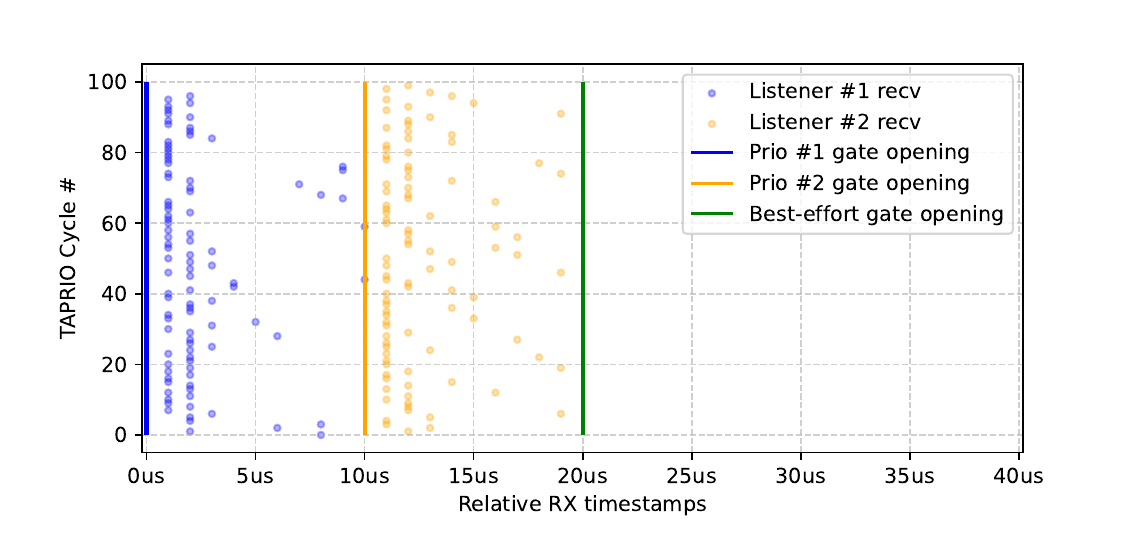}
		\caption{TSN metadata proxy enabled}
		\label{fig:proxy}
	\end{subfigure}
	\caption{RX timestamps mapped to 802.1Qbv (\texttt{taprio} Qdisc) cycles.
		Without the proxy (\ref{fig:noproxy}), the priority info lost and every packet uses the best-effort timeslot.
		With TSN proxy (\ref{fig:proxy}) the priorities preserved and the packets received within their timeslot.}
	\label{fig:operation}
\end{figure}

The test consisted of a simple single node Kubernetes environment.
We focused on integration and functionality, performance testing was not performed.
We used Ubuntu 24.04, 24.10 and Debian Trixie GNU/Linux distributions with kernel versions 6.8, 6.11 and 6.12 respectively.
The TSN proxy was implemented as a \textit{DaemonSet} and deployed with the \texttt{kubectl apply -f tsn-metadata-proxy.yaml} command.
We used two popular Kubernetes distributions, \textit{Minikube} (v1.34) and \textit{KIND} (v0.24).
As the main CNI plugin, we used Flannel, but we confirmed that the TSN proxy works with other CNI plugins as well, such as Kindnet and Calico.

Our test scenario had two TSA talkers running in pods, generating priority 1 and 2 packets.
We also had two listeners outside the node, where we capture the received packets with their timestamps.
On the node, we configured \texttt{taprio} Qdisc with 40 usec cycle time.
This is divided into 3 gates (timeslots): 0-10~$\mu$s prio 1, 10-20~$\mu$s prio 2, and 20-40~$\mu$s for best effort with prio 0 (which is the default for all packets).
If a packet reaches the \texttt{taprio} outside of it's cycle, it will be queued until it's gate opens.

Without the TSN proxy, the talker's priorities are deleted before they reach the NIC and the \texttt{taprio}.
As a result, they fall into the best-effort timeslot shown in Fig.~\ref{fig:noproxy}.
Note that packets received outside of their timeslot are transmitted immediately after their gate opens.
Therefore, we have more timestamps right after the gate opens than in the rest of the slot.
With TSN proxy (Fig.~\ref{fig:proxy}), the priorities of the TSA packets are preserved.
They are placed in their proper time slots and respect the correct gate openings.
As one can see in the figure, listener \#1 receives packets in timeslot prio 1 and listener \#2 receives packets in timeslot prio 2.

It is important to note that the gate configuration in this scenario may differ from a real TSN scheduling.
This scheduling is designed to make a point and help understanding.

\section{Conclusion}

We have designed and evaluated a TSN metadata proxy that allows unmodified TSAs to be deployed as microservices using Kubernetes.
The TSN proxy is implemented as a simple extension CNI plugin, which uses eBPF to store the metadata required for scheduling.
This CNI plugin is deployed as a companion to a primary CNI plugin, which handles the interface, address, routing, and policy configuration of the pods.
Finally, we compare our solution to other work and discuss the main differences.

\section*{Acknowledgement}
This work was supported by the European Union’s Horizon 2020 research and innovation programme through DETERMINISTIC6G project under Grant Agreement no. 101096504.
The authors would like to thank Dr. Balázs Varga, Dr. János Farkas, István Moldován, Dr. Miklós Máté, and Dr. János Harmatos for their help and insightful feedback.

\bibliographystyle{ieeetr}
\bibliography{reference}

\end{document}